\documentstyle[sprocl,epsfig]{article}

\def\Journal#1#2#3#4{{#1} {\bf #2}, #3 (#4)}

\def\PLB{{\em Phys. Lett.}  B}

\def\PRD{{\em Phys. Rev.} D}
\def\ZPC{{\em Z. Phys.} C}

\def\be{\begin{equation}}
\def\ee{\end{equation}}
\def\bea{\begin{eqnarray}}
\def\eea{\end{eqnarray}}

\begin{document}

\title{SPIN ASYMMETRIES IN PROTON-PROTON SCATTERING AT HIGH ENERGIES AND
       MODERATELY LARGE MOMENTUM TRANSFER}

\author{ S.V.Goloskokov}

\address{Bogoliubov Laboratory of
Theoretical
  Physics, Joint Institute for\\  Nuclear Research, Dubna
  141980, Moscow
  region, Russia.\\E-mail: goloskkv@thsun1.jinr.dubna.su}


\maketitle\abstracts{We study $pp$ scattering at high energies and 
moderately large
momentum transfer using a QCD--based model in which the proton is viewed
as being composed of a quark and a diquark. This
model leads to spin asymmetries which are
neither small nor vanish at high energies. The predicted
ratio of helicity flip and non-flip amplitudes is about
0.2-0.3 and
$A_n$ asymmetry is about 20-30\% for $|t| > 4 \,\mbox{GeV}^2$.
}

Theoretical investigation of spin effects in high-energy 
exclusive proces-ses at moderately large momentum transfer is one 
of the unsolved problems in QCD. In view of the polarization 
physics programs proposed for the future proton accelerators 
\cite{prop}, this problem is very important. Massless QCD leads 
to hadronic helicity conservation and to zero single-spin 
asymmetries. Mass  and higher order perturbative QCD corrections 
lead to non-vanishing single-spin transverse asymmetries but 
they are much smaller than the experimental results.

There are many experimental observations of large spin effects at 
high energies and moderately large momentum transfer \cite{hesp}. 
The low-energy results for the $A_n$ asymmetry   at $p_B= 
28\;{\rm GeV}$ in the BNL \cite{krish} are of the same order of 
magnitide as the FNAL observations \cite{fnalp} at $p_B= 
200\;{\rm GeV}$ and similar values of $t$. In spite of large 
experimental errors, the conclusion might be done  that spin 
effects in high-energy reactions exhibit a weak energy 
dependence.

In this report, we are interested in spin effects at high 
energies and moderately large momentum transfer ($3\;{\rm GeV}^2 
<|t| <<s$). Our approach is based on the diquark picture 
\cite{kroll} where the proton is viewed as being composed of a 
quark and a diquark in the dominant valence Fock state instead of 
three quarks 
\begin{equation}
\label{pwf}
|p,\lambda\rangle  = f_S\,\varphi_S(x_1)\,B_S\, u(p,\lambda)
             + f_V\, \varphi_V(x_1)\, B_V
              (\gamma^{\alpha}+p^{\alpha}/m)\gamma_5
              \,u(p,\lambda)/\sqrt{3}\, .
\end{equation}
The scalar(S) and vector(V) diquarks provide an effective 
description of non-perturbative effects. Their composite nature 
is taken into account by diquark form factors. The diquark 
picture of the proton simplifies our calculations drastically due 
to the reduced number of constituents. We use that model in 
combination with the two-gluon exchange picture as a 
representative of the Pomeron. 

In the kinematic region of interest, the double helicity--flip 
amplitudes are believed to be much smaller than the helicity 
non--flip ones and the two non-flip amplitudes are equal in
magnitude approximately. Therefore, we have to calculate or to 
model a non-flip and a flip amplitude only. We can, for 
convenience and without loss of generality, fix the helicities of 
two protons at $+1/2$. The structure of the amplitude then 
simplifies to 
\begin{equation} 
\label{t} 
T_{ \lambda_4  +; 
\lambda_2 +} = \bar u(p_4,\lambda_4)[s A(t,s) + \hat p_1 B(t,s)] 
u(p_2,\lambda_2) 
\end{equation} 
The two-gluon graphs for the colour--singlet $t$-channel exchange 
have been considered for the invariant function $A$. The function 
$A$ is calculated under the assumption that the t-channel gluons 
couple to one constituent, a quark or a diquark at the helicity 
non-flip vertex. Into the helicity flip vertex, we include the 
perturbative $\alpha_s$ correction. Hence, we consider minimally 
connected graphs which allow us to keep all constituents collinear. 
The used set of graphs leads to gauge-invariant scattering 
amplitudes. The invariant function $B$, dominating the helicity 
non-flip amplitudes, is parametrized by a standard 
phenomenological Pomeron. We use two different fits \cite{gol_kr} 
which qualitatively describes the differential cross section of 
elastic $pp$ scattering for $|t| \ge 3 \,\mbox{GeV}^2$.
\begin{figure}[htb]
   \vspace*{-.3cm}
   \hspace*{-1.3cm}
   \epsfysize=5cm\epsfxsize=6.3cm\epsffile{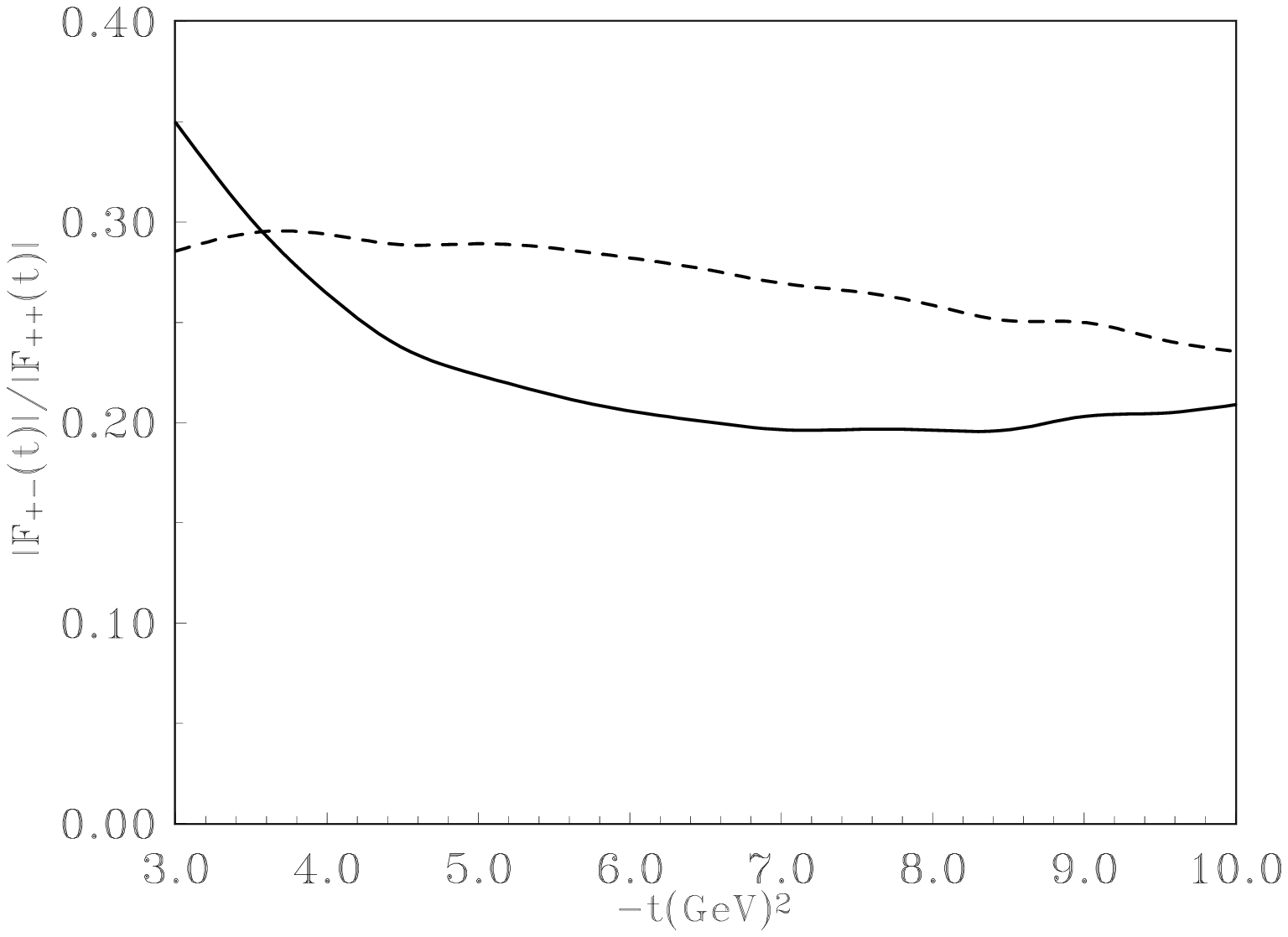}
   \hspace*{-2.8cm}
   \epsfysize=5cm\epsfxsize=6.3cm\epsffile{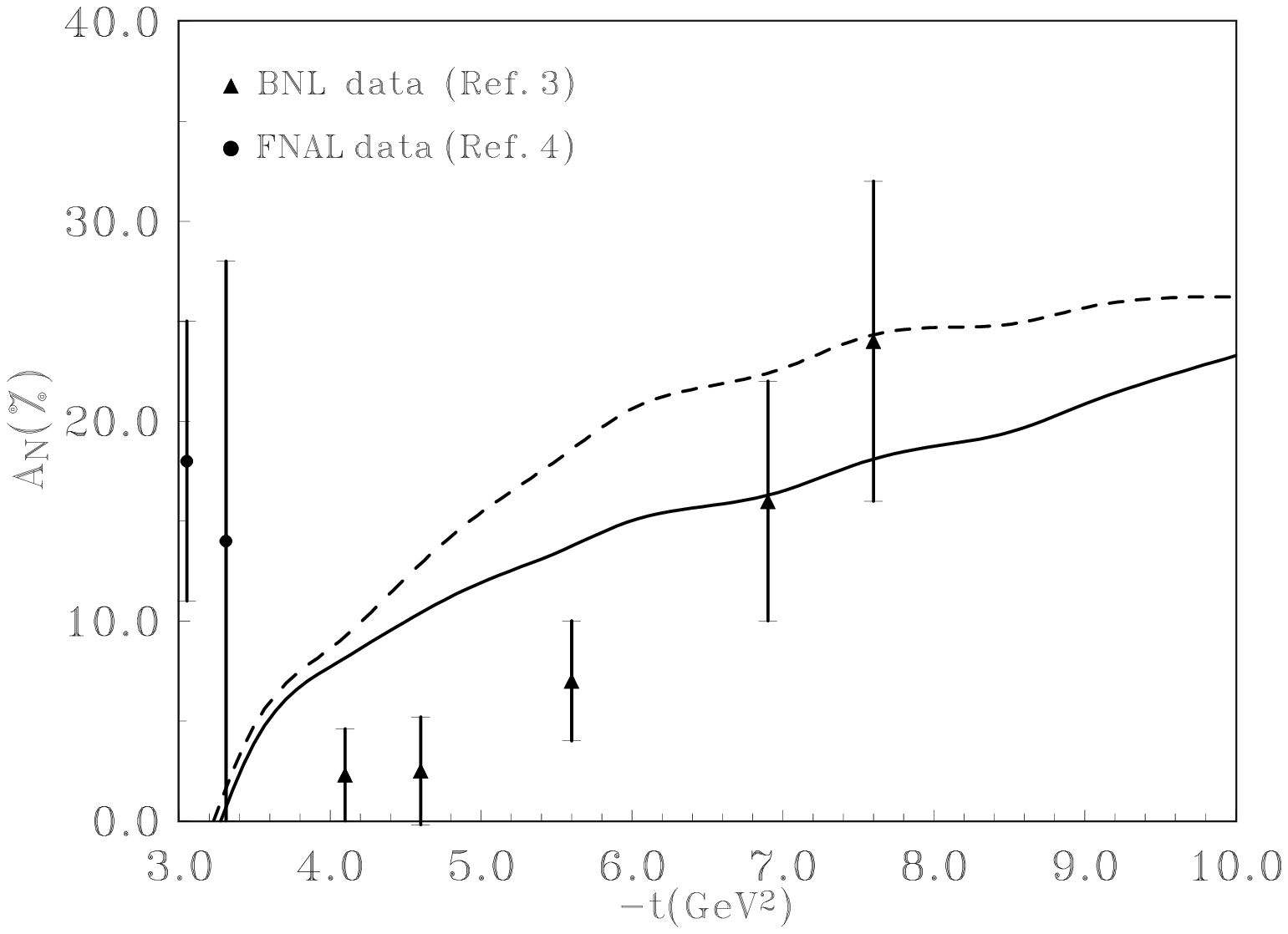}
   \vspace*{-0.3cm}
   \centerline{\hspace{3.5cm} Fig.1 \hspace*{4.8cm} Fig.2}
   \vspace*{-0.3cm}
\caption{ Absolute value of the ratio of helicity-flip to non-flip amplitudes.}
\vspace*{-0.1cm}
\caption{ Model predictions for single-spin asymmetry
 for two fits of the $B$ amplitude $^6$.}
\vspace*{-0.2cm}
\end{figure}

The absolute value of the ratio of $A$ to $B$ is proportional to 
the ratio of helicity-flip and non-flip amplitudes (see Fig.\ 1). 
This ratio  is fairly large $|A|/|B| \sim 0.1 \,\mbox{GeV}^{-1}$ at $|t| 
\ge 3\, \mbox{GeV}^2$ thus indicating a substantial amount of 
helicity flips generated through the vector diquarks in the 
model.

 In our model the helicity flips are generated by vector 
diquarks. It turns out that the invariant function $A$ is of 
substantial magnitude and not in phase with the Pomeron 
contribution. Our model, therefore, provides a single-spin 
asymmetry $A_N$ that is rather large for momenta transfer $|t| 
\ge 3\, \mbox{GeV}^2$. Our prediction for $A_N$ is shown in Fig.\ 
2. The predicted asymmetry amounts to about 20--30\% for $|t| > 6 
\,\mbox{GeV}^2$; it is of the same order of magnitude as has been 
observed in the BNL experiment \cite{krish} and the FNAL 
experiment \cite{fnalp}. The decrease of the asymmetry at smaller 
momenta transfer is connected with the observed zero of ${\rm 
Re} A$. The double spin transverse asymmetry in this kinematic 
region is rather large in our model \cite{gol_kr}. It turns out 
to be of an order of 10-15\% for  $|t| > 4 \,\mbox{GeV}^2$. Our 
results for  spin asymmetries are rather close to those 
obtained in \cite{gol-95} although the latter are valid in the 
momentum transfer region $2\,\mbox{GeV}^2 <|t| < 
4\,\mbox{GeV}^2$.

Thus, on the basis of the diquark model we have calculated  spin 
effects in high-energy proton-proton scattering at moderately 
large momentum transfer. The spin-flip effects in the model are a 
consequence of the quark-diquark structure of the proton, which 
reflects the non-perturbative physics in the hadronic binding. 
Spin-1 diquarks which appear besides spin-0 diquarks as 
constituents of protons can change their helicity when 
interacting with gluons. Besides the quark-diquark structure, our 
model is based on the t-channel exchange of a colour-singlet 
two--gluon system and, in so far, bears resemblance to the 
Pomeron exchange. The important feature of the spin effects 
obtained in the diquark model is their approximate energy 
independence. On the other hand, they decrease with increasing  
momentum transfer. Our results are valid at large $s$ and 
moderate momenta transfer, larger than  few GeV$^2$. This 
kinematical region can be investigated, for instance, in the 
proposed HERA-$\vec N$ experiment \cite{prop}.

I am grateful to P.Kroll for a fruitful collaboration. This work was supported in part by the Heisenberg-Landau Grant.

\end{document}